\documentclass[12pt]{article}
\pdfoutput=1

\usepackage{graphicx} 

\usepackage{enumerate}
\usepackage{hyperref}
\usepackage[retainorgcmds]{IEEEtrantools}
\usepackage[utf8]{inputenc}
\usepackage[justification=justified,font=small,labelfont=bf]{caption}
\usepackage{bbold}

\usepackage{graphicx}
\usepackage{dcolumn}
\usepackage{bm}
\usepackage{amsfonts}
\usepackage{amsthm}
\usepackage{amsmath}

\usepackage{amssymb}
\usepackage{epsfig}
\usepackage{color}
\usepackage{textcomp}

\def\be{\begin{equation}}
\def\ee{\end{equation}}
\def\ba{\begin{eqnarray}}
\def\ea{\end{eqnarray}}

\def\bdm{\begin{displaymath}}
\def\edm{\end{displaymath}}

\def\bq{\begin{quote}}
\def\eq{\end{quote}}

 at 10truept

\newcommand{\bea}{\begin{eqnarray}}
\newcommand{\eea}{\end{eqnarray}}

\newcommand{\bi}{\begin{itemize}}
\newcommand{\ei}{\end{itemize}}

\newcommand{\Mpl}{M_\mathrm{Pl}}
\newcommand{\St}{St\"uckelberg\ }

\newcommand{\beq}{\begin{equation}}
\newcommand{\eeq}{\end{equation}}
\newcommand{\beqa}{\begin{eqnarray}}
\newcommand{\eeqa}{\end{eqnarray}}

\def\ltap{\ \raise.3ex\hbox{$<$\kern-.75em\lower1ex\hbox{$\sim$}}\ }
\def\gtap{\ \raise.3ex\hbox{$>$\kern-.75em\lower1ex\hbox{$\sim$}}\ }
\def\gl{\ \raise.5ex\hbox{$>$}\kern-.8em\lower.5ex\hbox{$<$}\ }
\def\roughly#1{\raise.3ex\hbox{$#1$\kern-.75em\lower1ex\hbox{$\sim$}}}

\begin{document}

\thispagestyle{empty}
\begin{flushright}
August 2018
\end{flushright}
\vspace*{1.5cm}
\begin{center}
{\Large \bf de Sitter Branes  in a Flat Bulk of Massive Gravity}\\

\vspace*{1.7cm} {\large Nemanja Kaloper\footnote{\tt
kaloper@physics.ucdavis.edu} and James H.C. Scargill\footnote{\tt
jhcscargill@gmail.com}}\\

\vspace{.3cm} {\em Department of Physics, University of
California, Davis, CA 95616, USA}\\

\vspace{1.5cm} ABSTRACT
\end{center}
We construct de Sitter branes in a flat bulk of massive gravity in $5D$. We find two branches of 
solutions, reminiscent of the normal and self-accelerating branches in DGP, but with rather different properties.
Neither branch has a self-accelerating limit: the background geometry requires having 
a nonvanishing tension.  On the other hand, on both branches there are sub-branches where 
the leading order contributions of the tension to the curvature cancel. In these cases it turns out that larger tensions curve the background less. Further, both branches support a localized $4D$ massless graviton for a special choice of bulk mass terms. This choice may be protected by enhanced gauge symmetry. 
Finally, we generalize the solutions to the case of bigravity in a flat $5D$ bulk.

\vfill \setcounter{page}{0} \setcounter{footnote}{0}
\newpage

\section{Introduction}

Changing gravity in the far infrared emerged as an alternative for addressing the Universe's missing mass problem, which in General Relativity (GR) is modeled as either dark energy or dark matter. This program is fraught with difficulties\footnote{The modifications which never encounter such difficulties seem to be
merely different gauges of scalar-tensor gravity, possibly with irrelevant operator corrections. At this point these are clearly less interesting.} with predictivity of the theory in the regime between the cosmological and short-distance scales, which are particularly acute  when the modifications break or alter diffeomorphism invariance of GR, which is really a gauge redundancy of the theory. New degrees of freedom
may appear lending to all kinds of trouble. A notorious problem are the ghost modes---the excitations with negative kinetic terms---that are  commonly encountered on generic backgrounds \cite{deser,Higuchi:1986py,nima}. It now appears that the ghosts are not a showstopper. Frameworks were found where the gauge redundancies are relaxed to allow full massive spin-2 multiplets, yet without propagating ghosts \cite{drgt,fawad}. 

These models have been applied as approximations to GR in the hope that interesting alternatives to
missing mass can be found. If the gravitational field is massive but very light, one might hope that the modifications
of the gravitational dynamics, be it at the level of the background or perturbations, could feature new phenomena
that can alter cosmological dynamics and compact object fields in interesting ways \cite{guido,mukoh,Comelli:2011zm,koyama}. However these applications run into a problem: the helicity-0 mode of the massive graviton tends to be strongly coupled and its sector runs out of control at very low scales. For example, one may take the mass of the graviton to be $m \sim H_0 \sim 10^{-33} \, {\rm eV}$, in order to defer the deviations from GR to horizon scales, where the dark energy component of the missing mass kicks in. In this case, in vacuum the breakdown happens at typically $\sim 1000  \, {\rm km}$ \cite{nicorata}. In denser backgrounds, this scale can be shorter, as has been noted in \cite{clare}, but it is still very low. These arguments have subsequently been refined and 
confirmed\footnote{There is an increasing effort underway to address some cosmological problems in massive gravity cosmology by  bigravity models.}
by various authors \cite{saffin,gaba,javi}. 

In response it has been proposed that the theory can be improved by embedding massive gravity in warped extra dimensions. It is well known that some of the problems with perturbativity of massive spin-2 modes are relieved in AdS spaces \cite{kogan,maximus,karch}. Therefore it is fathomable to entreat that similar setups can improve massive gravity too \cite{gaba}. In turn, if such frameworks are to be used in phenomenological applications, the extra dimensions either need to be compactified, or a theory of matter needs to be confined to a brane that floats in
the higher-dimensional bulk. In the latter case the brane theory needs to be properly covariantized in the massive gravity bulk. This means that terms which add to the usual Gibbons-Hawking action on the boundary are required to
ensure that the theory has a well defined action principle and Hamiltonian evolution. In the case of flat bulks this has
been considered in an interesting article \cite{Gabadadze:2018bpf}, which summarizes with a call for deployment of the ``machinery" of \cite{sasaki} to study cosmology of such braneworlds and in particular their vacua,
given by the geodesic worldvolumes of an empty, but possibly tensional brane in the bulk. 

This ``machinery", i.e. the formalism of the Gauss-Codazzi-Weingarten equations that yield the tool for determining the intrinsic geometry of a hypersurface in the bulk, is fortunately not necessary when the vacua are maximally symmetric subspaces. Their large symmetry translates into relatively
simple embedding equations \cite{weinbergbook}, that provide a shortcut for constructing solutions. Such methods have been employed in the case of bent braneworlds in flat and curved bulks \cite{kaloperlinde,bent}. In the case of massive gravity the same shortcut remains available, with the generalizations of the boundary conditions (a.k.a. Israel junction conditions) outlined in \cite{Gabadadze:2018bpf}.

Following this route, in this paper we construct the vacua of an orbifold brane in a flat bulk of massive gravity \cite{drgt} and bigravity \cite{fawad}, where the intrinsic geometry on the brane is de Sitter. We work with a $5D$ bulk and $4D$ brane for simplicity. We impose orbifolding, i.e. identification of the different sides of the brane which serves as a $\mathbb{Z}_2$ mirror. We then compute the relationship between the expansion rate, the tension and the graviton mass with both finite and infinite bulk boundary conditions, the two branches of solutions corresponding to the so called normal and self-accelerating branches in DGP gravity \cite{dgp} (see, e.g. \cite{specters,gkmp} for a review).  We  find interesting differences between these solutions and the vacua found by similar means in the DGP braneworlds \cite{deffayet}. In particular, the vacua on both branches never feature the self-accelerating limits: the tension can never be zero.

However, both branches feature a completely novel behavior for large tension. On the branch where the bulk extends to infinity, when tension is negative, there is a sub-branch where the larger the tension, the smaller the expansion rate on the brane! On the other branch the same happens for positive tension. 
In other words, the leading order effects of the tension cancel out from the induced curvature. This is reminiscent 
of degravitation \cite{degrav} and the mechanism of vacuum energy sequester \cite{sequester}. 

We also find that on both branches of solutions there  exists a normalizable $4D$ graviton localized on the brane. This occurs for a very special selection of the bulk mass parameters. The reason is that due to the degeneracy of the bulk mass contributions, the mass of this mode 
can be tuned to zero, similarly to \cite{peloso,El-Menoufi:2014aza}. These cases cannot be realized as a continuous limit of a massive theory, because of the Higuchi bound \cite{Higuchi:1986py}. 

The localized graviton is the zero mode of the bulk equation, which means that the tensor sector of the theory
does not contain ghosts. In fact, since our starting point is ghost-free massive gravity, this is not surprising: the
only possible source of a ghost would be the brane boundary, which might introduce it via the coupling 
of the brane bending mode with gravity \cite{dgp,specters} in the scalar sector, just like in DGP. While we did not check this explicitly it is likely that since there are two branches of background solutions, the scalar ghost could be
absent on at least one of them because its bulk wavefunction is not normalizable. In this case the full nonlinear theory
would be ghost-free, and since it contains the massless $4D$ graviton, there would be an enhanced gauge symmetry protecting the special values of the parameters. This is just the usual $4D$ diffeomorphism invariance of GR. A small deviation of the parameters from the special value yields a ghost, which suggests that the perturbation theory will not correct the special value, in order to maintain unitarity. In this case the additional helicities of the localized graviton decouple, since their couplings to matter sources are $\propto m = 0$. Finally we extend the construction of the background to the case of bigravity.

The solutions which we provide look very interesting, since the cancellation of the tension contributions from the
brane intrinsic curvature may be helpful to attempts for addressing the cosmological constant problem in this setup. At the same time, finding the localized massless $4D$ graviton, and raising the bulk graviton mass ought to improve the range of validity of the low energy theory. However, before these appealing features can be put to a good use, it
is necessary to verify that the theory can accommodate consistent low energy dynamics, including $4D$ cosmology. We leave these important questions for the future.

\section{Boundary Terms in Massive Gravity} \label{sec-boundary terms}

We begin by briefly reviewing the emergence of the brane terms required to covariantize the bulk masses in spacetimes with a boundary. As \cite{Gabadadze:2018bpf}  we use the same form of the bulk action, 
concentrating on doubly flat bulks. The boundary terms which one obtains now come both from the Einstein-Hilbert action and from the graviton mass terms. The variation of the Einstein-Hilbert action yields total derivatives which
vanish in the absence of the brane, but in its presence leave a non-zero projection on the brane given by the extrinsic curvature, which needs to be cancelled to ensure correct Hamiltonian evolution. This yields the Gibbons-Hawking extrinsic curvature term.

In the presence of the bulk graviton mass terms additional terms
are necessary. The reason is that when the fiducial metric is fixed by the background, covariantizing the theory
requires introduction of \St fields, whose variation formally changes the mass terms. Now, these changes can be undone by coordinate transformations in the bulk, and thus they are symmetries. Since the theory is covariant, thanks to the presence of \St fields, this means that the variation of  \St fields alone, or the diffeomorphisms which compensate them, must add up to gauge transformations: they must sum up to total derivatives, which do not alter the action. This happens thanks to the determinant form of the mass terms, such as in the flat space example like 
\be
\int_M d^Dx \, \frac{1}{(D-n)!} \epsilon^{\mu_1 \cdots \mu_n \lambda_1 \cdots \lambda_{D-n}} \epsilon_{\nu_1 \cdots \nu_n \lambda_1 \cdots \lambda_{D-n}}  \, \partial^{\nu_1} \partial_{\mu_1} \pi \cdots \partial^{\nu_n} \partial_{\mu_n} \pi \, , 
\ee
where $\pi$ is really absorbed into the helicity-0 mode of the massive graviton in the axial gauge. While this is a total derivative in the bulk, a boundary at a finite location will change this since this term mixes with the brane bending mode, and yields a brane-localized contribution. To preserve the original invariance, it is this term that needs to be cancelled by adding brane-localized terms which absorb it away. 

In the case of a $5D$ bulk and a codimension-1 orbifold brane, the appropriate action which includes the boundary terms\footnote{When the brane is a $Z_2$ orbifold, the boundary action should be divided by $2$ when we neglect the mirror image of the bulk. We will assume this wherever necessary.} is \cite{Gabadadze:2018bpf}
\ba \label{full action}
S &=& \frac{M_5^3}{2} \int_M d^5X \sqrt{-g} \left[ R - \frac{m^2}{4} \sum_{n=2}^5 \alpha_n \epsilon \epsilon \left( \mathbb{1} - \sqrt{g^{-1}f} \right)^n \mathbb{1}^{5-n} \right] \nonumber \\
&+& M_5^3 \int_{\partial M} d^4x \sqrt{-\gamma} \left[ K + \frac{m^2}{4} \Sigma(\Phi) \sum_{n=1}^4 \alpha_{n+1} \epsilon^{(4)} \epsilon^{(4)} \left( \mathbb{1} - \sqrt{\gamma^{-1}\phi} \right)^n \mathbb{1}^{4-n} \right] + \cdots \, .
\ea
The ellipsis denotes brane-localized matter contributions, including the tension. We use the shorthand notation 
$$
\epsilon \epsilon A^n B^{D-n} = \epsilon^{\mu_1 \cdots \mu_n \rho_1 \cdots \rho_{D-n}} \epsilon_{\nu_1 \cdots \nu_n \lambda_1 \cdots \lambda_{D-n}} A^{\nu_1}_{\mu_1} \cdots A^{\nu_n}_{\mu_n} B^{\lambda_1}_{\rho_1} \cdots B^{\lambda_{D-n}}_{\rho_{D-n}} \, .
$$
Our convention for the Levi-Civita's\footnote{$\epsilon_{0123\ldots} = 1, \, \epsilon^{0123\ldots} = -1$. It appears the literature is not uniformized on this.} is the standard $4D$ Poincare algebra convention, reflecting the sign flip due to raising indices with $\eta^{\mu\nu}$.  The signs in (\ref{full action}), with the condition that $\alpha_2  > 0$, guarantee that the graviton mass squared is nonnegative. 
Here $\epsilon^{(4)}$ is the pullback of $\epsilon$ to boundary. The induced metric on the boundary is $\gamma_{\mu\nu}$ and its extrinsic curvature is $K_{\mu\nu}$. The term $\Sigma(\Phi)$ is what we now need to determine.

The auxiliary metric in arbitrary coordinates is $f_{AB} = \partial_A \Phi^I \partial_B  \Phi^J \eta_{IJ}$, where $\Phi^I$ are the Cartesian \St fields and their pullback to the brane is $\phi_{\mu\nu} = \partial_\mu \Phi^I \partial_\nu \Phi^J P^K_{\phantom{K}I} P^L_{\phantom{L}J} \eta_{KL}$.\footnote{Note that all indices, not just the free ones, must be projected onto the brane.} The brane coordinates $x^\mu$ are independent combinations of the bulk coordinates $X^A$ constrained by the brane worldvolume equation 
$\Omega(\Phi) = 0$. The projection matrix acting on the auxiliary metric is given by $P^I_{\phantom{I}J} = \delta^I_J - n^I n_J$, where the normal to the brane $n_A$ is defined by $d \Omega = \partial_A \Omega dX^A = n_A dX^A$. This follows since a displacement of a point on the brane
$X^A$ to a nearby point $\tilde X^A$ defines a parallel transport along the brane that must be orthogonal to the normal, which therefore must be proportional to the gradient of $\Omega$ at $\Omega=0$. 
Note that the projection matrix has one zero eigenvalue, directed along the normal $n^I$, which follows due to the brane
location constraint $\Omega(\Phi)=0$. Similar equations can be written relating the dynamical metrics $g_{AB}$ and $\gamma_{\mu\nu}$. Note that this implies that we will be employing the ans\"atz $f_{AB} = g_{AB}$ in the construction of the background metric.\footnote{This is a calculational simplification which does not change the generality of our analysis. The graviton mass terms in (\ref{full action}) can be rewritten as $\sum_n \beta_n \epsilon\epsilon (\sqrt{g^{-1} f})^n \mathbb{1}^{5-n}$ where $\beta_n$ are linear combinations of $\alpha_n$, including the `cosmological constant' terms for $g$ and $f$. To allow a flat bulk, the former must be tuned by hand, and the latter can then be selected accordingly to ensure the existence of a flat bulk. Changing the ans\"atz to, e.g, $f_{AB} = c^2 g_{AB}$ merely picks different linear combinations of $\beta_n$ which contribute to $\alpha_n$. 
We stress that our choice to pick flat bulks is a tuning. By itself, this doesn't `solve' or `unsolve' the cosmological constant problem. Since the only aspect of the cosmological constant problem in quantum field theory coupled to (semiclassical) gravity is the UV-(in)sensitivity of its terminal value, as discussed in, 
e.g., \cite{sequester,selftuning}, the choice we make is merely a hidden sector tuning. As long as one is only interested in the effects of brane-localized sources, such as the brane vacuum energy, this tuning does not matter in the semi-classical gravity limit \cite{selftuning}. To go beyond this limit one would need the full UV 
completion of massive gravity, which is unavailable at present. }

We are interested in embedding a bent de Sitter brane in a $5D$ Minkowski bulk which for simplicity we can describe with flat Cartesian coordinates $Z^I$, and the Cartesian Minkowski metric 
\be
ds^2_5 = \eta_{IJ} dZ^I dZ^J \, .
\ee
Now we can embed the brane by using a standard construction explained in \cite{weinbergbook}, which amounts to
imposing a constraint
\be
\Omega = \eta_{IJ} Z^I Z^J- \frac{1}{H^2} = 0\, .
\label{constraint}
\ee
To understand this intuitively, Wick-rotate the background flat geometry to Euclidean signature. Since Euclidean de Sitter is a sphere, we need to embed a spherical brane in flat $5D$ space. To accomplish this, all we need to do is pick a sphere of a given radius $1/H$, cut the space along it and specify the boundary conditions on the sphere.  Since we restrict to a $\mathbb{Z}_2$ orbifold, we remove either the interior or exterior, and replace the other side of the sphere by the mirror image of the side we wish to retain. Clearly, the brane is the boundary of the cut and a discontinuity. The discontinuity means that the brane must carry an action, which includes stress energy, to cause the gravitational field flux lines to jump and to maintain the diffeomorphism invariance of the underlying theory.

Next, to determine the boundary terms $\sim \Sigma(\Phi)$, we first determine the normal to the brane boundary by using the brane worldvolume equation (\ref{constraint}). From the definition above the components of the normal $n_I$ are given by
\be
d\Omega = n_I dZ^I |_{\Omega=0}  \, .
\ee
We can normalize the normal to unity, bearing in mind our convention for its direction: the normal is oriented away from the brane, into the bulk, on either side of the brane, so as to reflect the ${\mathbb Z}_2$ symmetry. We will give the explicit form of the normal shortly after we pick the explicit form of the intrinsic coordinates to the brane. 

Now we consider the diffeomorphism invariance in the presence of the brane. As we noted, the idea is to transform either the \St fields or the bulk coordinates, but not both, to find how the brane terms arise. Since the bulk is flat, and the brane is at a fixed location, for simplicity we can just translate the bulk coordinates while keeping the \St fields fixed. We use the translation 
\be
Z^I \rightarrow \tilde Z^I = Z^I - \partial^I \pi \, , 
\label{transl}
\ee
which does not change the fiducial metric, but only the dynamical bulk metric. Thus
\be
\tilde f_{IJ} = f_{IJ} = \eta_{IJ}\, , ~~~~~~~~~~ \tilde g_{IJ} = g_{KL} \frac{\partial Z^K}{\partial \tilde Z^I} \frac{\partial Z^L}{\partial \tilde Z^J} = \eta_{KL} \Bigl(\delta^K_I + \partial_I \partial^K \pi \Bigr) \Bigl(\delta^L_J + \partial_J \partial^L \pi \Bigr) \, . 
\ee
Thus, inverting the metric $\tilde g_{IJ}$,\footnote{$\tilde g^{IJ} = \eta^{KL} \Bigl(\delta_K^I - \partial^I \partial_K \pi \Bigr) \Bigl(\delta_L^J - \partial^J \partial_L \pi \Bigr)$, notice the sign flip.} yields
\be \label{varf}
\delta^I_J - \sqrt{ \tilde g^{IK}\tilde f_{KJ}} = \nabla^I \nabla_J \pi  \, ,
\ee
where we have trivially replaced the partial derivatives with covariant ones on the right hand side, since 
we are using Cartesian coordinates in the bulk. This is a technical trick which will come in handy shortly. 

Finally, substituting (\ref{varf}) in the bulk graviton potential term in (\ref{full action})  and keeping the variation of the action 
$\propto \nabla \nabla \pi$ yields terms like 
\ba \label{bulk td term}
- \int d^5Z \sqrt{-g} \epsilon^{I \cdots}\epsilon_{J \cdots} \nabla^J \nabla_I \pi \cdots &=& -
\int d^5Z \sqrt{-g} \nabla^J \left( \nabla_I \pi \epsilon^{I K \cdots} \epsilon_{J L \cdots} \nabla^L \nabla_K \pi \cdots \right)   \nonumber \\
&=&  \int_{\partial \mathcal{M}} d^4 x \sqrt{-\gamma} n^I \nabla_I \pi \, n^J n_K \epsilon^{KL \cdots} \epsilon_{J N \cdots} \nabla^N \nabla_L \pi  \, .
\ea
The first step follows from the determinant structure of the mass terms, and the rest from Gauss' theorem. 
The last line follows from the projection of the embedding coordinates $Z$ onto the boundary brane, whose intrinsic 
coordinates are $x$. In the braneless bulk this term vanishes by Gauss' law applied to the whole spacetime, with appropriate boundary conditions at infinity. All the tensors in the last term should now be restricted to the brane coordinates $x^\mu$ which realize a cover of $\Omega = 0$, except for
$n_I \partial^I \pi = n^I \partial_I \pi = \partial_4 \pi$ where the derivative here is with respect to the coordinate running along the unit normal to the brane. Thus in the presence of the brane, the boundary projection of the bulk term (\ref{bulk td term}) is nonzero,  
\be \label{bulk td term1}
\int_{\Omega = \{0^+\} \cup \, \{ 0^-\}} d^4x \,  \partial_4 \pi \epsilon^{4 \mu\cdots} \epsilon_{4 \nu\cdots} \nabla^\nu \nabla_\mu \pi \cdots \, .
\ee
If left in the theory, this term would break translational symmetry on the brane, and violate energy conservation. To properly covariantize the theory with a brane, it must be cancelled by a term on the brane. 

Since we take the translation (\ref{transl}) to move both the bulk and the brane, in the active sense, this means that in the new coordinates the brane is inserted at
\be
\eta_{IJ} \tilde Z^I \tilde Z^J = \frac{1}{H^2} \, .
\ee
We define the translations in this way in order to ensure that the intrinsic geometry of the brane remains unaffected by a bulk translation. 
Thus the location of the brane is changed by the Cartesian vector $\tilde Z^I - Z^I = \partial^I \pi$ projected onto the normal $n_I$. Rewriting this as
\be
n_I \tilde Z^I - n_I Z^I =  n_I \delta Z^I = - n_I \partial^I \pi = - \partial_4 \pi
\ee
we can interpret the coefficient of the brane variation $n_I \partial^I \pi$ as the variation of $ -n_I Z^I$. To cancel the bulk variation (\ref{bulk td term}), we add to the action the brane term as displayed in (\ref{full action}) with a function
\be
\Sigma = n_I Z^I \, .
\label{sigma}
\ee
Let's demonstrate this. First to define the brane action we project the induced fiducial metric on the brane. Since before the translation we have $\phi_{\mu\nu} = \gamma_{\mu\nu}$, projecting the shifted fiducial bulk metric onto the brane we find 
\be \label{phi mu nu} 
\tilde \phi_{\mu\nu} = \nabla_\mu \tilde{Z}^I \nabla_\nu \tilde{Z}^J P^K_{\phantom{K}I} P^L_{\phantom{L}J} \eta_{KL} = \gamma_{\mu\lambda} \left( \delta^\lambda_\rho - \nabla^\lambda \nabla_\rho \pi \right) \left( \delta^\rho_\nu - \nabla^\rho \nabla_\nu \pi \right) \, ,
\ee
which immediately implies 
\be
\delta^\mu_\nu - \sqrt{\gamma^{\mu\lambda} \tilde \phi_{\lambda\nu}} = \nabla^\mu\nabla_\nu \pi  \, . 
\ee
Thus we can rewrite the variation of the bulk action projected on the brane, with our signs convention from (\ref{full action}), as
\begin{equation}
\int_{\partial \mathcal{M}} d^4 z \, \partial_4 \pi  \epsilon^{ \nu \cdots} \epsilon_{\mu \cdots} \left( \delta^\mu_\nu - \sqrt{\gamma^{\mu\lambda} \tilde \phi_{\lambda\nu}} \right) \cdots \, . 
\end{equation}
Therefore we need to add to the brane a term whose variation is the exact opposite, so the two cancel out. 
From our conventions in (\ref{full action}) and the discussion above, this clearly means that we should choose $\Sigma$ as given in Eq. (\ref{sigma}). Note that we can formally covariantize this term by recalling that the Cartesian \St fields are $\Phi^I = Z^I$ and so \cite{Gabadadze:2018bpf}
\begin{equation}
\Sigma(\Phi) = n_I \Phi^I = \Phi^4 \,  . \label{sigma constant coordinate}
\end{equation}
In fact our calculations above could all have been carried out using \St fields and their variations, which is a more convenient method for  general backgrounds. 

At this point we are ready to write down the full set of the metric field equations. 
The metric field equations coming from \eqref{full action} in the bulk and on the boundary are
\ba  \label{bulk eom} 
&&G_{AB} -  T_{AB}/ M_5^{3}  =  \\
&&~~~~~~~~~~~~~~~ - \frac{m^2}{8} \sum_{n=2}^5 \alpha_n \left( \epsilon \epsilon \left( \mathbb{1} - \sqrt{g^{-1}f} \right)^{n-1} \mathbb{1}^{5-n} \right)^C_{\phantom{C}D} \nonumber \\
&&~~~~~~~~~~~~~~~~~~~~~~~~~~~~~~~~~~~~~~~~~~~~~ \times \left( n\, \delta^D_{(A} g_{B)E} \sqrt{g^{-1} f}^E_{\phantom{E}C} + \left( \mathbb{1} - \sqrt{g^{-1} f} \right)^D_{\phantom{D}C} g_{AB} \right) \, , \nonumber 
\ea
and 
\ba \label{boundary eom}
&&K_{\mu\nu} - K \gamma_{\mu\nu} - S_{\mu\nu}/M_5^{3} =  \\
&&~~~~~~~~~~~~~~~~~ \frac{m^2}{4} \Sigma(\Phi) \sum_{n=1}^4 \alpha_{n+1} \left( \epsilon^{(4)} \epsilon^{(4)} \left( \mathbb{1} - \sqrt{\gamma^{-1}\phi} \right)^{n-1} \mathbb{1}^{4-n} \right)^\lambda_{\phantom{\lambda}\rho}  \nonumber \\
&&~~~~~~~~~~~~~~~~~~~~~~~~~~~~~~~~~~~~~~~~~~~~~~~~~~~~  \times
\left( n \, \delta^\rho_{(\mu} \gamma_{\nu)\sigma} \sqrt{\gamma^{-1} \phi}^\sigma_{\phantom{\sigma}\lambda} + \left( \mathbb{1} - \sqrt{\gamma^{-1}\phi} \right)^\rho_{\phantom{\rho}\lambda} \gamma_{\mu\nu} \right)  \, ,
\nonumber
\ea
where $S_{\mu\nu}$ is the brane borne stress energy. The boundary term $\Sigma$ is given by (\ref{sigma constant coordinate}). 

With all this set in place, we can
now turn to determining the vacuum solutions. We set the brane stress-energy to be pure tension $\sigma$ 
and set any other matter sources to zero. This gives us a very useful shortcut to 
determining the solution. We do not need to solve the equations directly; we know that the bulk equations are 
trivially solved.
%
%
 All that remains is to write down the explicit relationship of the embedding coordinates of $4D$ de Sitter in flat $5D$ bulk \cite{weinbergbook}, which satisfy (\ref{constraint}), and interpret
the $5D$ coordinates of the embedding as the \St fields. This will yield the relationship between the intrinsic curvature length $H$, the tension $\sigma$ and the bulk graviton mass $m$.

\section{de Sitter branes} \label{sec-dS brane}

\begin{figure*}[thb]
\centering
\includegraphics[scale=0.3]{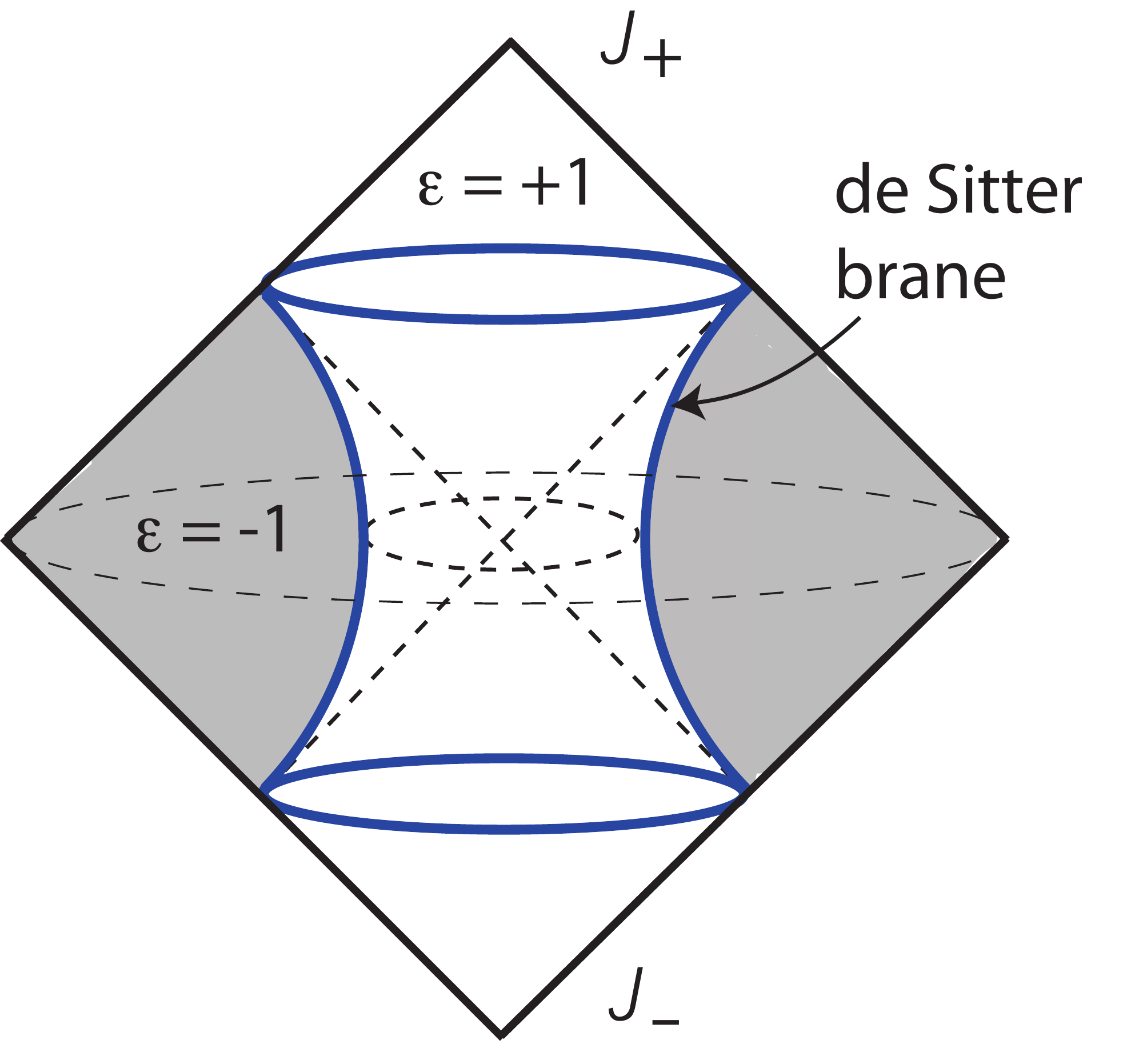}
\caption{dS brane in flat bulk. We pick the unshaded interior when $\varepsilon H>0$, and the shaded exterior when $\varepsilon H < 0$. These choices define two different branches of solutions.}
\label{fig2}
\end{figure*}

An explicit coordinate realization of the constraint (\ref{constraint}) which transforms the Cartesian Minkowski coordinates to the brane-intrinsic coordinates $x^\mu$ and the orthogonal coordinate $x^4 =w$ is  \cite{weinbergbook}
\ba \label{embedding}
Z^0 &=& (1 - \varepsilon H w ) \left( \frac{e^{-Ht}}{4 H} \left( -1 + e^{2Ht} H^2 |\mathbf{x}|^2 \right) + \frac{e^{Ht}}{H} \right) \, , \nonumber \\
Z^i &=& (1 - \varepsilon H w ) \,  x^i \,  e^{Ht} \, , \\
Z^4 &=& (1 - \varepsilon H w ) \left( \frac{e^{-Ht}}{4 H} \left( -1 + e^{2Ht} H^2 |\mathbf{x}|^2 \right) - \frac{e^{Ht}}{H} \right) \, . \nonumber
\ea
The constraint equation (\ref{constraint}) which fixes the location of the brane becomes
\be
w = 0 \, .
\ee
Substituting the coordinate transformations (\ref{embedding}) into the bulk metric $ds_5^2 = \eta_{IJ} dZ^I dZ^J$  
yields the metric describing the bulk ending on the $\mathbb{Z}_2$ brane at $w=0$,
\begin{equation}
ds^2 = \Bigl(1 - \varepsilon H |w| \Bigr)^2 \Bigl( - dt^2 + e^{2Ht} d\mathbf{x}^2 \Bigr) + dw^2 \, .  \label{flat with dS brane}
\end{equation}
Here $H$ is the intrinsic curvature radius of the brane geometry, and $ \varepsilon = \pm 1$ controls `which side' of the brane at $w = 0$ is retained (for a positive $H$). Away from the cut hypersurface $w=0$ the geometry (\ref{flat with dS brane}) is just flat space. The Penrose diagram of the geometry of this embedding is given in Fig. (\ref{fig2}). 

What remains to determine is the relationship between the curvature radius $1/H$, brane tension $\sigma$, and the scale which controls the IR modification of gravity, in this case the bulk graviton mass $m$. To this end we can turn to the equation (\ref{boundary eom}), since the bulk equation (\ref{bulk eom}) is trivially satisfied away from the brane cut. Since on the
background solution $\mathbb{1} - \sqrt{\gamma^{-1}\phi} = 0$ and $S_{\mu\nu} = - \sigma \gamma_{\mu\nu}/2$, the extrinsic curvature equation (\ref{boundary eom})
reduces to
\be
K_{\mu\nu} - K \gamma_{\mu\nu} = - \frac{\sigma}{2M_5^3} \gamma_{\mu\nu} - \frac{3}{2} \alpha_2 m^2 \Sigma \gamma_{\mu\nu} \, . \label{junction}
\ee
We recall that $\alpha_2$ is degenerate with the overall normalization of the graviton mass terms. Thus we can absorb it away, choosing $\alpha_2 = \frac{2}{3}$ from now onwards, which ensures that $m$ sets the numerical value of the graviton mass. Since the normal to the brane in any of the cases is $\partial_w$ as can be checked from the embedding map (\ref{embedding}), using \eqref{sigma constant coordinate} for $\Sigma$ we have
\begin{equation}
\Sigma = \frac12 \partial_w Z^I \eta_{IJ} Z^J \big|_{w=0^+} = 
 \frac{1}{2} \partial_w \left( (H^{-1} - \varepsilon w)^2 \right) \big|_{w=0^+} = -\frac{1}{\varepsilon H} \, ,
\end{equation}
and correspondingly for its mirror image on the other side of the brane. Finally for the metric \eqref{flat with dS brane}, the extrinsic curvature of the $w=0^+$ hypersurface is $K_{\mu\nu} = n_A \Gamma^A_{\mu\nu} = \varepsilon H \gamma_{\mu\nu}$, and correspondingly for $0^-$. Therefore the junction condition \eqref{junction} yields 
\be
3 \varepsilon H = \frac{\sigma}{2M_5^3}  - \frac{m^2}{\varepsilon H} \, . 
\label{backh}
\ee
Curiously, this equation falls in the class of phenomenological suggestions for background curvature radii equations in which IR modifications of gravity might affect our determination of dark energy \cite{dvaliturner}. 

One interesting feature of this equation is that the last term obstructs taking the limit $H \rightarrow 0$, which would 
not happen if $m=0$. Indeed, if one takes a de Sitter brane in a flat bulk of $5D$ Einstein gravity \cite{kaloperlinde}, one
can readily take the limit $H \rightarrow 0$ by sending $\sigma$ to zero. It is interesting to understand the physical reason for this obstruction. Without the graviton mass, the embedding (\ref{embedding}) requires the rescaling of 
$z^I$ coordinates to keep them regular. 

This of course corresponds to the standard picture of how one flattens a sphere: pick a point, say the North Pole, take a disk around it of some fixed radius smaller than the radius of the sphere, and then simultaneously send both radii to infinity. Mathematically, this corresponds to the Wigner-Inonu
contraction of the de Sitter group to Poincare. However in the presence of the graviton mass the scaling symmetry is 
broken at $H=0$---e.g. by the dimensional \St fields, which on the background are equal to the embedding coordinates. Hence the singular behavior as $H \rightarrow 0$. 

To determine the explicit form of the branches of solutions, we rewrite (\ref{backh}) as a standard quadratic equation, which since $\varepsilon^2 = 1$ is
\be
H^2 - \frac{\varepsilon \sigma}{6M_5^3} H  + \frac{m^2}{3} = 0 \, . 
\label{backhh}
\ee
The roots of this equation are
\be
H = \frac{\varepsilon \sigma}{12 M_5^3} \pm \sqrt{\frac{\sigma^2}{144 M_5^6} - \frac{m^2}{3}} \, ,
\label{branches}
\ee
where $\varepsilon = \pm 1$ shows which region of the bulk is retained with the brane as per Fig. (\ref{fig2}). 
For $\varepsilon = +1$, where in the bulk we retain the interior of the ``hourglass" surface in (\ref{fig2}), the solution
yields $H$ which is strictly positive for positive tension:
\be
H = \frac{\sigma}{12 M_5^3} \pm \sqrt{\frac{\sigma^2}{144 M_5^6} - \frac{m^2}{3}} \, .
\ee
If we start with the sub-branch with the positive sign of the root, $H$ starts very large for large positive tension and reaches the minimum value for $\sigma_* = 48 M_5^3 m$
where the square root vanishes, and so $H = \sigma_*/12 M_5^3$. There, the solution transitions to another sub-branch where the sign in front of the root flips, and $H$ decreases to zero, which again resides in the limit $\sigma \rightarrow \infty$. Throughout $H$ and $\sigma$ remain non-negative. Moreover $\sigma \ge \sigma_* = 48 M_5^3 m$ is gapped from zero, since the term in the square root $\propto m^2$ is negative.

On the other branch of solutions, where $\varepsilon = -1$ we keep the exterior of the ``hourglass" surface in (\ref{fig2}). Since this requires $H \ge 0$  (this can be viewed as a `gauge fixing' since the solutions with $H<0$ are readily mapped to these by $w \leftrightarrow -w$, $t \leftrightarrow -t$), when $\varepsilon = -1$, the tension is negative, $\sigma <0$. Again, $H$ starts from large values when $\sigma \rightarrow - \infty$ on the sub-branch with positive square root, decreases to $|\sigma_*|/12 M_5^3$ which occurs when $\sigma = - |\sigma_*| = -48 M_5^3 m$ and continues to decrease as $\sigma$ rebounds back to $-\infty$. 

It is now interesting to highlight some features of the solutions and differences relative to the previous examples such as
dS branes in flat Einstein bulks and DGP (for reviews see \cite{kaloperlinde} and \cite{specters}):
\begin{itemize}
\item in (\ref{branches}) the roles of the tension and the IR modification scale (here the graviton mass) are interchanged relative to the example of DGP, and the sign of the term outside of the square root is flipped;
\item as noted, once the sign of $\varepsilon$ is picked, $H$ varies continuously from infinity to zero on each branch, as a function of the tension; however, for each value of tension there are {\it two} values of $H$;
\item the $\varepsilon = +1$ branch, where the interior in Fig (\ref{fig2}) is retained, with finite volume, consistently reproduces the geometry of an inflating wall in flat Einstein bulk when $m=0$, $H = \frac{\sigma}{6 M_5^3}$ \cite{kaloperlinde}; on the branch $\varepsilon = -1$, this happens when the tension is negative; 
\item however, the sign of the $\propto m^2$ term is {\it negative}; the graviton mass term cancels against the tension's contribution to the intrinsic curvature on the brane, similarly to the negative cosmological constant in cutoff AdS braneworlds \cite{bent};
\item as a result, on both branches there is no ``self-acceleration": the tension $\sigma$ is gapped from zero, and thus ``self-acceleration"---positive intrinsic curvature even when tension is zero---is impossible;
\item however there is now a completely novel feature, which is very interesting, and which appears on both branches: when $\varepsilon = +1$, on the sub-branch with negative square root, in the limit of large positive tension, $\sigma \gg m M_5^3$, the tension cancels to leading order from $H$; expanding the square root we find
$H \simeq \frac{\sigma}{12 M_5^3} - \frac{\sigma}{12 M_5^3} \Bigl( 1 - \frac{24 m^2 M_5^6}{ \sigma^2} \Bigr)
\simeq 2 \frac{m^2 M_5^3}{ \sigma}$; on the other branch, $\varepsilon = -1$, the same happens for negative tension. 
\end{itemize}
This last limit of the solutions yields the equation
\be
H \simeq 2 \frac{m^2 M_5^3}{|\sigma|} \, , 
\label{selfg}
\ee
which is unusual. It shows features of degravitation \cite{degrav} and vacuum energy sequester \cite{sequester}. Here, the larger the tension of the brane, \underbar{the less it curves the space}! The curvature 
of the vacuum is see-sawed from the vacuum energy of the field theory which inhabits the brane (as measured by the tension) by a `pivot scale' $\sim m^2 M_5^3$. Given how curious this is, it makes sense to look at just how much this might help with the cosmological constant problem in this setup---i.e. how large a hierarchy could be reliably induced between the vacuum curvature $H$ and the vacuum energy $\sigma$ in a controllable regime of field theory.

The cutoff of the theory is given by the lowest strong coupling scale in the theory, which is set by the
helicity-0 bulk graviton, and is given by $\Lambda_{7/2} \sim (M_5^{3/2} m^2)^{2/7}$ \cite{gaba}. For the cancellation which led to (\ref{selfg}) to make sense all terms should be within the range of the effective field theory, i.e. below the cutoff. This means that the largest magnitude of the tension, which radiative corrections would saturate, and the calculation would not be outside of the regime of validity of perturbation theory, would be
$|\sigma_*| \sim 12 M_5^3 \Lambda_{7/2}$. Thus the natural vacuum value of $H$ would be
\be
H_* \sim \frac{1}{6} \frac{m^2}{\Lambda_{7/2}} \sim \frac{1}{6} \Bigl( \frac{m}{M_5} \Bigr)^{3/7} m \, .
\label{Hsigma}
\ee
Taking the bulk graviton mass to be e.g. $m \sim 10^{-3} \, {\rm eV}$, which could yield interesting corrections
to sub-millimeter gravitational forces, and taking $M_5$ to be as low as it could be for bulks which
behave as sub-millimeter compact spaces, $M_5 \sim 10^{17} \, {\rm eV}$, one gets
$H \sim 10^{-13} \, {\rm eV}$. 

While remarkable, this reduction of the vacuum curvature falls short of the mark by nearly twenty orders of magnitude, setting the radius of the universe to be of the size of the Solar system, roughly. Nevertheless, the reduction from the naively expected `natural' value of $\sigma/M_5^3 \sim \Lambda_{7/2} \sim {\rm MeV}$, a reduction of some $19$ orders of magnitude, is quite curious. This feature of the solutions might warrant more scrutiny. Of course, in order for it to have any aspiration to help with the real world, 
the theory would need to produce another spin-2 mode to start with, much lighter than $10^{-3} \, {\rm eV}$, to have a working chance to approximate GR. 
Further, in such a setup one would need to verify that as the tension is varied the 
curvature of the universe $H$ retains the scaling with tension as in (\ref{Hsigma}) for degravitation to work. 
Note that 
this is the physical observable which determines the background geometry of the vacuum, irrespective of the
Planck mass which physically only controls the mutual interaction of probes of this geometry, such as, e.g., the deflection of their trajectories. 

In the next section we will outline a very simple realization of such a scenario, where the bulk localizes an exactly 
massless $4D$ graviton on the brane, which may be able to support the $4D$ FRW cosmology with corrections due to the massive bulk modes. We will determine the `local' Planck mass due to the zero mode and find that 
it is independent of the brane tension in the degravitating limit.

\section{Localization of $4D$ gravity on the brane}

In light of the discussion above, a theory with a de Sitter brane in a flat bulk might improve the chances 
of massive gravity to address the cosmological constant problem if it had an additional light spin-2 field. This field 
could approximate GR at large distances, while the massive graviton might control the background and remove the
brane-localized vacuum energy from the brane intrinsic curvature. A natural way to realize an extra light mode is to
have it emerge from the bulk by localizing it to the brane, as for example in the RS2 case \cite{RS2}. It could happen
that the localized spin-2 could turn out to be much lighter, even massless, for a special set of graviton mass parameters and given brane boundary conditions, as in \cite{peloso,El-Menoufi:2014aza}.

We will show that similar phenomena occur here. We will work from the start with spin-2 $4D$ metric fluctuations, which are traceless, transverse and massless in the $4D$ sense, propagating with a pole at $p^2 = 0$.
Thus we use the perturbation of (\ref{flat with dS brane})  
\begin{equation}
ds^2 = \Bigl(1 - \varepsilon H |w| \Bigr)^2 ds_{4\,dS}^2  + h_{\mu\nu} dx^\mu dx^\nu + dw^2 \, , \label{flatdSpert}
\end{equation}
where the metric perturbation satisfies traceless-transverse conditions, $h = {\cal D}_\mu h^\mu{}_\nu = 0$ and is $4D$ massless, 
$({\cal D}^2 - 2 H^2) h_{\mu\nu} = 0$, where ${\cal D}_\mu$ are $4D$ de Sitter covariant derivatives. 

Off the brane the field equations are just given by \eqref{bulk eom}. Linearizing them in
$h_{\mu\nu}$ and using our gauge conditions, with our choice $\alpha_2 = 2/3$,
\be
\nabla_A \nabla^A  h_{\mu\nu} - 2 \nabla^A \nabla_{(\mu} h_{\nu) A} - m^2 h_{\mu\nu}  = 0 \, . \label{linearizedbulk}
\ee
Commuting the covariant derivatives in the second term in (\ref{linearizedbulk}), separating out the $w$-derivatives and  extracting the warp factor dependence yields 
$ \nabla_A \nabla^A h_{\mu\nu}  -2 \nabla^A \nabla_{(\mu} h_{\nu) A} =\partial_w^2 h_{\mu\nu} + 
 ({\cal D}_\mu {\cal D}^\mu  - 2 H^2)  h_{\mu\nu}/(1-\varepsilon Hw)^2 =   \partial_w^2 h_{\mu\nu}$, since the last two terms after the second equality cancel due to $h_{\mu\nu}$ having zero $4D$ mass. Separating now the bulk wavefunction from the 
$4D$ de Sitter helicity tensor for a traceless transverse spin 2 mode, $h_{\mu\nu} = \Psi \varepsilon_{\mu\nu}$,
\be
\Big( \partial_w^2 - m^2 \Big) \Psi = 0 \, .
\ee
The bulk wavefunctions of the graviton zero modes are linear combinations of decaying and growing exponentials, modulated by the bulk graviton mass:
\be
\Psi = A e^{mw} + B e^{-mw} \, ,
\ee
as long as these modes can satisfy the boundary conditions on and off the brane. 

To find the boundary condition on the brane we follow the procedure outlined in
\cite{specters}, and perturb the junction conditions (\ref{boundary eom}). 
After straightforward algebra we find (see, for example, \cite{specters,garta} for the general procedure)
\ba
\delta \Bigl( K_{\mu\nu} - K \gamma_{\mu\nu} \Bigr) &=& -\frac12 \partial_w h_{\mu\nu} - 4 \varepsilon H h_{\mu\nu} \, ,\nonumber \\
\delta \Bigl(\frac{S_{\mu\nu}}{M_5^3} \Bigr)&=& - \frac{\sigma }{2 M_5^3} h_{\mu\nu} \, ,  \\ 
\delta \theta_{\mu\nu} &=&  \frac{m^2}{2\varepsilon H} \Bigl(1 - \alpha_3 \Bigr) h_{\mu\nu} \, , \nonumber 
\ea
where we have used $\theta_{\mu\nu}$ to denote the right hand side of Eq. (\ref{boundary eom}) and as before set $\alpha_2 = 2/3$. Combining these terms, eliminating the brane tension using the background equation (\ref{backh}), and factoring out the $4D$ helicity tensor yields the boundary condition for the localized wave function,
\be
\partial_w \Psi(0) = -2 \varepsilon H \Bigl[ 1 - \frac{m^2}{2H^2} \bigl(1+ \alpha_3 \bigr) \Bigr] \Psi(0) \, .
\label{bcpsi}
\ee

Now, on the $\varepsilon = -1$ branch, $w \in [0,\infty)$. Thus for the graviton bulk wavefunction to be normalizable, it should be 
\be
\Psi = e^{-mw} \, , 
\ee
and so $\partial_w \Psi(0) = - m\Psi(0)$. Substituting into (\ref{bcpsi}) along with $\varepsilon = -1$ gives an equation which determines the value of $\alpha_3$ required to have the localized normalizable 
$4D$ graviton zero-mode:
\be
\alpha_3 = -\Bigl( 1 +  \frac{H}{m} - 2\frac{H^2}{m^2} \Bigr) \, .
\label{locneg}
\ee
When $H/m \ll 1$, this means that the parameter $\alpha_3$ is just slightly smaller than $-1$. Note that in this case the tension is negative, and hence the brane bending mode may have a negative kinetic term. By itself this is not a definitive sign of the presence of a ghost, however one would have to check the full spectrum of perturbations, not only traceless transverse tensors, to ensure the theory remains healthy. 

On the other hand, in the case $\varepsilon = +1$, the variable $w$ is in a compact interval. It satisfies
$w \in [0,1/H]$. Because of this, in general both $\exp(\pm mw)$ can contribute, and in addition to the boundary condition on the brane (\ref{bcpsi}) we need another boundary condition on the horizon. There, we take the Dirichlet boundary condition, $\Psi(1/H) = 0$. This generalizes the boundary condition for the normalizable mode on the 
de Sitter brane in $5D$ flat bulk of Einstein gravity. It takes into account that the horizon is not the end of space, but 
merely a causal boundary, and that wavepackets can cross it without any reflection (in infinite proper time of a distant observer). In contrast any other boundary condition, for example Neumann or mixed, would produce nonzero reflection, obstructing traversability of horizons, and implying presence of hard null boundaries bearing localized and highly blue-shifted energy density.

Hence the zero mode radial wave function for $\varepsilon = +1$ is 
\be
\Psi = \sinh\Bigl[\frac{m}{H}(1-Hw)\Bigr] \, .
\ee
Since $\partial_w \Psi(0) = - m \cosh(m/H)$ and $\Psi(0) =  \sinh(m/H)$, substituting into (\ref{bcpsi}) we find that now
$\alpha_3$ must be
\be
\alpha_3 = - \Bigl(1 - \frac{H}{m} \coth(\frac{m}{H}) - 2 \frac{H^2}{m^2} \Bigr). 
\label{alpha3pos}
\ee
In the limit where the tension cancels out we again can have $H \ll m$,  and so $\alpha_3$ is just a tiny bit larger than $-1$, since $\coth(m/H) \rightarrow 1$.

Note that in both cases, the boundary value problem which the radial wave function solves is self-adjoint, and that the zero mode wavefunction in each case has no nodes. This means that the zero mode wavefunction is a minimum energy state of the tower of traceless transverse tensors, which therefore does not contain any unstable modes. Moreover, since the lightest traceless transverse tensor is $4D$ massless, it contains only two helicity-2 modes. The helicity-0 and helicity-1 modes decouple in this limit, as can be readily verified in perturbation theory \cite{beng}. 

Further, our initial starting point is ghost-free massive gravity in five dimensions. If any instabilities were to arise with the inclusion of a brane, they would only arise as instabilities of the boundary, i.e. IR problems. The one and only new gravitational degree of freedom which the brane brings in is the brane bending mode, which mixes with gravity. This mode does not come with a tower of states, and can be a scalar ghost \cite{specters}. The same happens in the usual braneworld models in which the bulk theory is GR, and the dimensionally reduced effective action contains a tower of massive gravitons. However on some of the branches of solutions the bulk wavefunction of the ghost is not normalizable, and so the ghost decouples from the dynamics in the full nonlinear theory. Given that our construction involves two branches of solutions, one with a finite and the other with an infinite bulk volume, it is very likely that the same happens here, although we have not checked it directly. 

Thus if the brane bending mode is not normalizable, the effective $4D$ theory on that branch will be completely ghost-free at the full nonlinear level. The nonlinear terms needed to complete the linear theory are completely fixed since they all arise by dimensional reduction from the action (\ref{full action}) which would be ghost free on this branch. Thus this branch would have the short multiplet of spin 2 as the lightest mode, realizing precisely the $4D$ diffeomorphism invariance of standard General Relativity. The absence of the ghost would therefore imply that the gauge symmetry of the theory is enhanced in the full nonlinear effective $4D$ theory, singling out the special values of $\alpha_3$. This enhanced gauge symmetry would be the operational mechanism in $4D$, protecting the ghost-free branch from all corrections, perturbative and nonperturbative, which cannot break the symmetry. Note again, that since the initial bulk theory is ghost free, all it takes to realize this is to make the brane bending mode non-normalizable. Exciting such modes would require breaking the gauge symmetry with infinitely strong couplings, completely invalidating the breaking mechanism. 

A diagnostic of the protection mechanism in perturbation theory is that if $\alpha_3$ deviates ever so slightly from the two special values, the lightest mode will be massive, with a very small mass inside the Higuchi forbidden window \cite{Higuchi:1986py} (regardless of the sign of $m^2$). This would bring in a ghost in the lightest mode (helicity-0 or helicity-1). 
However as long as there is a ghost-free branch, this mustn't happen, implying that such deformations cannot be generated in perturbation theory with the special value of $\alpha_3$ due to a strong coupling `barrier', which a healthy theory can never cross by unitarity. 

Thus when the brane bending ghost is not normalizable, including the non-linear interactions of the effective brane graviton cannot---by itself---introduce any fundamental ghosts. Nonlinearities will not break the gauge symmetries of the tensor sector as they all arise from the ghost-free gravity in the bulk by dimensional reduction. In fact we expect that in this case the effective $4D$ theory would be a structure resembling bigravity, with a zero mode $4D$ graviton, a lightest KK state of massive gravity playing the role of the massive graviton in bigravity, plus an additional tower of heavier KK states. Again, we have not checked this explicitly. Such a check would be very interesting.

Returning to the special values of $\alpha_3$, it is interesting to also consider what the mass of the next lightest graviton is. For $\varepsilon = -1$ there will be a continuum of states, with mass gap $m$, whereas for $\varepsilon = +1$, there will be a discrete spectrum, and in this case one can deduce that the mass gap is at least $\sqrt{m^2 + \pi^2 H^2}$, and hence the Higuchi bound is not violated.

Since for both branches the zero mode gravitons are normalizable
we can compute the $4D$ Planck scale controlling their coupling to brane probes. 
In each case, the coupling of graviton fluctuations to the brane stress-energy is
\begin{equation}
\frac{1}{M_5^{3/2}} \int d^4x dw \sqrt{-g}\, h_{\mu\nu}(x,w) \, \delta(w)  \, S^{\mu\nu} = \frac{\mathcal{N}^{-1/2} \Psi(0)}{M_5^{3/2}} \int d^4x \sqrt{-\gamma}\, {\varepsilon}_{\mu\nu}(x) S^{\mu\nu} \, , 
\end{equation}
and so the effective four dimensional Planck mass is 
\be
\Mpl^2 = \frac{M_5^3 \mathcal{N}}{\Psi(0)^2} \, ,
\label{planckmass}
\ee
where $\mathcal{N}$ is the normalization factor of the graviton wavefunction in the bulk.

On the $\varepsilon = -1$  branch the norm of the radial wavefunction $\Psi$ is 
\be 
\mathcal{N} = 2 \int_0^\infty dw\, (1+Hw)^4 e^{-2m w} = \frac{1}{m} \Bigl( 1 + {\cal O} \bigl(\frac{H}{m} \bigr)\Bigr) \, . 
\ee
On the $\varepsilon = +1$ branch for $H \gtrsim m$ and the integral for ${\cal N}$ yields 
$M_{Pl}^2 \simeq M_5^3/H$, reproducing the behavior familiar from the case of de Sitter branes in $4D$ GR bulks \cite{kaloperlinde}. In the degravitating limit, $H/m \ll 1$ one finds
\be
\mathcal{N} = 2 \int_0^{H^{-1}} dw\, (1-Hw)^4 \sinh^2 \left( \frac{m}{H}(1-Hw) \right) 
\simeq  \frac{e^{2m/H}}{2m} \Bigl(1 +  \ldots \Bigr) 
\, .
\ee
In this case the exponential contribution to ${\cal N}$ will cancel against $\Psi(0)^2$ in the denominator in (\ref{planckmass}), and so we find that for the 
localized zero mode gravitons on both branches $M_{Pl}^2 \simeq M_5^3/m$ when $H \ll m$. In fact, this precisely reproduces the conditions of our numerical estimate for $H_*$ from the previous section. This offers opportunities for developing interesting phenomenology, however one should first verify that the theory admits limits that allow conventional $4D$ cosmology
to take place.

Moreover, we stress again that here we have only analyzed the traceless-transverse tensors. We have not considered scalar perturbations in the theory, which may also be normalizable. In fact it is known that such modes can lead to instabilities, as for example the ghost on the self-accelerating branch of DGP \cite{lpr,specters,koyamaghost}. A full perturbation analysis including modes other than just traceless transverse tensors would therefore seem to be warranted.

\section{Bigravity}

Another means to bring GR into the story is to start with bigravity, which contains the massless graviton from the start \cite{fawad}. This theory starts with massive gravity \cite{drgt}, and adds dynamics for the fiducial metric $f_{IJ}$. For the solutions which shield the tension from curvature, aside from the limit which 
supports the massless $4D$ graviton which is normalizable, this may be the only recourse. Here we produce the
solutions, but without a detailed dynamical analysis. 

Having promoted the fiducial metric to a new dynamical metric $f_{IJ}$, the action \eqref{full action} is then modified by adding a bulk kinetic term $\frac{M_f^3}{2} \int_M d^5Z \sqrt{-f} R(f)$, along with the corresponding Gibbons-Hawking boundary term on the brane, $M_f^3 \int_{\partial M} d^4x \sqrt{-\phi} K(\phi)$. The induced metric on the boundary is again $\phi_{\mu\nu} = \partial_\mu Z^I \partial_\nu Z^J P^K_{\phantom{K}I} P^L_{\phantom{L}J} f_{KL}$. For completeness we must also include an $f$-space brane tension $- \int_{\partial M} d^4 x \sqrt{-\phi} \sigma_\phi$. The field equations (\ref{bulk eom})  and (\ref{boundary eom}) are now supplemented with two additional equations,  
\be \label{feq}
G(f)_{IJ} =  \frac{\sqrt{-g}}{\sqrt{-f}} \frac{m^2}{8} \sum_{n=2}^5 n \alpha_n \left( \epsilon \epsilon \left( \mathbb{1} - \sqrt{g^{-1}f} \right)^{n-1} \mathbb{1}^{5-n} \right)^K_{\phantom{K}L} \sqrt{g^{-1} f}^L_{\phantom{L}N} \delta^N_{(I} f_{J)K}  \,,
\ee
and
\ba \label{phieq}
&& K(\phi)_{\mu\nu} - K(\phi) \phi_{\mu\nu} + \frac{\sigma_\phi}{2 M_f^3} \phi_{\mu\nu} = \\
&& ~~~~~~ - \frac{\sqrt{-\gamma}}{\sqrt{-\phi}} \frac{m^2}{4} \Sigma(\Phi) \sum_{n=1}^4 n \alpha_{n+1} \left( \epsilon^{(4)} \epsilon^{(4)} \left( \mathbb{1} - \sqrt{\gamma^{-1}\phi} \right)^{n-1} \mathbb{1}^{4-n} \right)^\lambda_{\phantom{\lambda}\rho} \sqrt{\gamma^{-1} \phi}^\rho_{\phantom{\rho}\sigma} \delta^\sigma_{(\mu} \phi_{\nu)\lambda} \, . \nonumber
\ea
Note the opposite sign of the boundary contributions $\propto m^2$ in Eq. (\ref{phieq}) relative to (\ref{boundary eom}), which follows from the fact that the boundary action depends on $\gamma^{-1} \phi$. Moreover, the variation with respect to $\phi_{\mu\nu}$ has a slightly different form than in (\ref{boundary eom}) since the 
brane action measure is independent of $\sqrt{-\phi}$.
  
The bulk equation can be solved by $f_{IJ} = \eta_{IJ}$. At the background level on the brane, our construction is simple with $\phi_{\mu\nu} = \gamma_{\mu\nu}$, which cancels many of the contributions from the graviton potential terms. Hence $K(f)_{\mu\nu} = \varepsilon H \phi_{\mu\nu}$. Using $\Sigma = \frac{-1}{\varepsilon H}$ and $\alpha_2 = 2/3$ as before, the boundary equation for the $\phi$ metric yields
\begin{equation}
3 \varepsilon H  = \frac{\sigma_\phi}{2 M_f^3} + \frac{m^2}{\varepsilon H}  \, .
\end{equation}
Note the different sign in front of the term  $\propto m^2$ compared to the equation (\ref{backh}) coming from varying $\gamma_{\mu\nu}$, Thus in order for these to be consistent with $\gamma_{\mu\nu} = \phi_{\mu\nu}$ 
we must choose the tension $\sigma_\phi$ to satisfy 
\begin{equation}
\frac{\sigma_\phi}{M_f^3} = \frac{\sigma}{M_5^3} - 4\frac{m^2}{\varepsilon H} = \pm \varepsilon \sqrt{\frac{\sigma^2}{M_5^6} - 48 m^2} \, , \label{bigravity tension relation}
\end{equation}
where the sign in front of the square root corresponds to that in \eqref{branches}. This completes the construction.
With this relation between the brane tensions, the expression for the curvature of the brane in terms of the $g$-space tension is identical to the massive gravity case; if \eqref{bigravity tension relation} fails to hold, then the construction is invalid, and a de Sitter brane cannot be supported in a flat bulk. The investigation of the full space of bigravity solutions is clearly interesting, but is beyond the scope of the present work.

\section{Summary}

Our construction of de Sitter branes in a $5D$ flat bulk of massive gravity reveals several unusual and interesting features. Let us outline them here:
\begin{itemize}
\item First, the background solutions do not display the phenomenon of self-acceleration: the brane geometry is de Sitter only for tensions larger in absolute value than a critical value $48 M_5^3 m$; the graviton mass does not drive the `repulsion' required to induce cosmic acceleration. 
\item Second, on each branch of solutions there is a sub-branch where the tension $\sigma$ (which measures the vacuum energy of a brane-resident QFT) is effectively shielded from curvature: the leading order contribution cancels, leaving only a correction which gives $H \sim 1/|\sigma|$ and a larger tension gravitates less. For the $\varepsilon = +1$ branch this happens for positive tensions, and for the other branch this is true of negative tension, though that may introduce ghost-like instabilities.
\item
Third, on both branches there is a localized $4D$ massless graviton mode, with only helicity-2 excitations. This requires a special choice of the bulk graviton mass, and may look like fine tuning at first sight. However, the presence of only two propagating helicities indicates that there may be the enhanced gauge symmetry for this mode in $4D$. Indeed, since the bulk theory is ghost-free, if the brane bending mode does not introduce a ghost, the full nonlinear $4D$ effective theory would be ghost free, and contain the zero mode, implying the presence of the same gauge symmetry as in GR. This is circumstantially supported by the fact that the special value of the bulk graviton mass results in the $4D$ graviton mode right in the middle of the Higuchi window, so that infinitesimal changes of the bulk mass would immediately yield ghosts and break gauge symmetry. Radiative corrections don't do this. 
\item Fourth, our analysis is performed only at the linear level, with traceless transverse spin-2 modes. 
It may happen that scalars introduce instabilities, which we have not studied here, that undermine the de Sitter brane stability and obstruct this argument. A direct check of what happens would be very interesting.
\item Fifth, in both cases of localized $4D$ massless gravitons, when $H/m \ll 1$ their coupling to brane probes is controlled by
$M_{Pl}^2 \sim M_5^3/m$. For $M_5 \sim 10^{17} \, {\rm eV}$ and $m \lesssim 10^{-3} \, {\rm eV}$ this might open the
road towards a rich gravitational phenomenology if there is a ghost free sector, with signatures which may be within reach of tabletop searches \cite{adelberger}. 
\item Sixth, it remains to be seen if the constructions which we have found really are good initial points for developing a new thrust into the phenomenological applications of modified gravity. It should be checked that 
our backgrounds are not infected with scalar ghosts; that they support normal $4D$ cosmology at low energies and late times; and that their low energy phenomenology is within bounds.
\item Finally, even if some of the problems emerge, there remain two obvious roads for corrective actions: a) bigravity, and we have provided the de Sitter brane solutions for this framework as well, and b) warping the bulk, and considering massive gravity in AdS spaces \cite{gaba}, by importing the bent brane solutions from RS2 \cite{bent}.
\end{itemize}
We hope to return to some of these issues shortly.

\vskip.75cm

{\bf Acknowledgments}: 
NK would like to thank CERN Theory Division for kind hospitality in 
the course of this work. This work is supported in part by the DOE Grants DE-SC0009999 and DE-SC0019081.

\end{document}